\documentclass[aps,amssymb,prl, amsmath, twocolumn, floatfix]{revtex4}
\usepackage{graphics}
\usepackage{psfrag}
\usepackage{epsfig}
\usepackage{subfigure}

\usepackage[dvips,usenames]{color}

\begin{document}
\bibliographystyle{prsty}

\title{The jamming transition and new percolation universality classes in particulate systems with attraction}
\author{Gregg Lois}
\author{Jerzy Blawzdziewicz}
\author{Corey S. O'Hern}
\affiliation{
Department of Mechanical Engineering, Department of Physics, Yale University, New Haven, Connecticut 06520-8284
}

\begin{abstract}
We numerically study the jamming transition in particulate systems
with attraction by investigating their mechanical response at zero
temperature.  We find three regimes of mechanical behavior separated
by two critical transitions---connectivity and rigidity percolation.
The transitions belong to different universality classes than their lattice
counterparts, due to force balance constraints. 
We also find that
these transitions are unchanged at low temperatures and resemble
gelation transitions in experiments on colloidal and silica gels.
\end{abstract}
\maketitle

The provocative conjecture that mechanical response at zero
temperature is linked to slow dynamics at non-zero temperature in
repulsive glassy systems has sparked tremendous interest in the
jamming transition at ``point J''~\cite{liunagel}.  At point J
athermal, frictionless systems with finite-range repulsive
interactions~\cite{zerotemprep} undergo a single critical transition
from an unjammed state with a vanishing static shear modulus to a
jammed state with a non-zero static shear modulus as
density is increased.  At the transition, it is observed that the
average number of particles participating in the connected network
jumps discontinuously from $0$ to $L^d$ (where $d$ is the spatial
dimension and $L$ is the system size), which makes the transition
first-order in the network order parameter.  Recent work has indeed
shown that properties of the jammed state are closely related to the
slow dynamics in highly compressed repulsive
glasses~\cite{dynfromjam}.

How does the jamming transition change for particulate systems with
attraction?  It is likely that attractive interactions will
qualitatively change the nature of the jamming transition.  For
example, thermal systems with attraction can form repulsive glasses,
but they can also form gels and attractive glasses~\cite{attglass}.
Moreover, at zero temperature the mechanical properties of attractive
granular materials and powders are quite different than
those for dry granular media with purely repulsive
interactions~\cite{athermalatt}.  While the jamming phase diagram of
Ref.~\cite{liunagel} needs revision to include attraction, the close
correspondence between mechanical properties at zero temperature and
dynamics at non-zero temperature is robust.  This has been
demonstrated in experiments on attractive colloidal
suspensions~\cite{weitz} where, as expected from the jamming phase
diagram, gelation occurs upon increasing density, decreasing
thermalization, and decreasing stress.

In this Letter, we explore the jamming transition in attractive
particulate systems by studying their mechanical response at zero
temperature.  
A central conclusion from our work is that 
repulsive jamming is fundamentally different than
attractive jamming, even for an infinitesimal amount
of attraction.  
Instead of a single first-order transition in the
purely repulsive systems, we observe two second-order
transitions in the attractive systems---connectivity and rigidity percolation.  
These two
transitions separate three distinct types of mechanical response and
exhibit critical exponents that differ from those measured in the
connectivity and rigidity percolation transitions without force balance
constraints.  
The transitions we observe at zero temperature are also present at
small but finite temperatures and resemble gelation.

\begin{figure}
\begin{center}
\begin{tabular}{c}
{\bf (a)} \\
\mbox{
\psfrag{L}{\Huge{$\ell$}}
\psfrag{r}{\Huge{$r_{ij}/\sigma_{ij}$}}
\psfrag{s}{\Huge{$1$}}
\psfrag{F}{\Huge{$F(r_{ij})$}}
\psfrag{fr}{\Huge{$\phi_\mathrm{R}$}}
\psfrag{fp}{\Huge{$\phi_\mathrm{P}$}}
\psfrag{xl}{\Huge{$\phi$}}
\psfrag{yl}{\Huge{$2PV/YC\langle \sigma_{ij}\rangle$}}
\scalebox{0.32}{\includegraphics{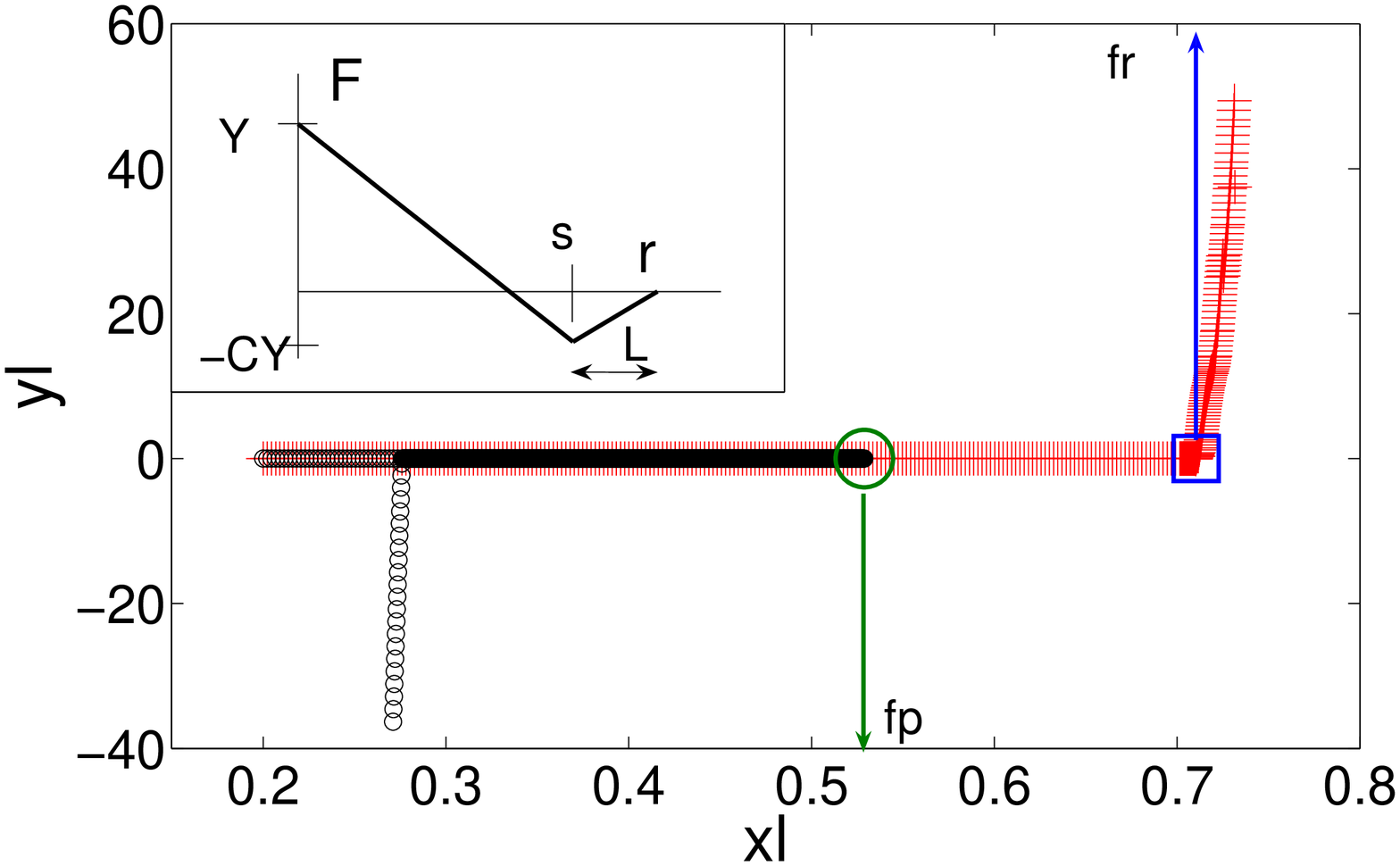}}
}\\
\end{tabular}
\newline
\newline
\begin{tabular}{cc}
{\bf (b)} & {\bf (c)} \\
\mbox{
\scalebox{0.16}{\includegraphics{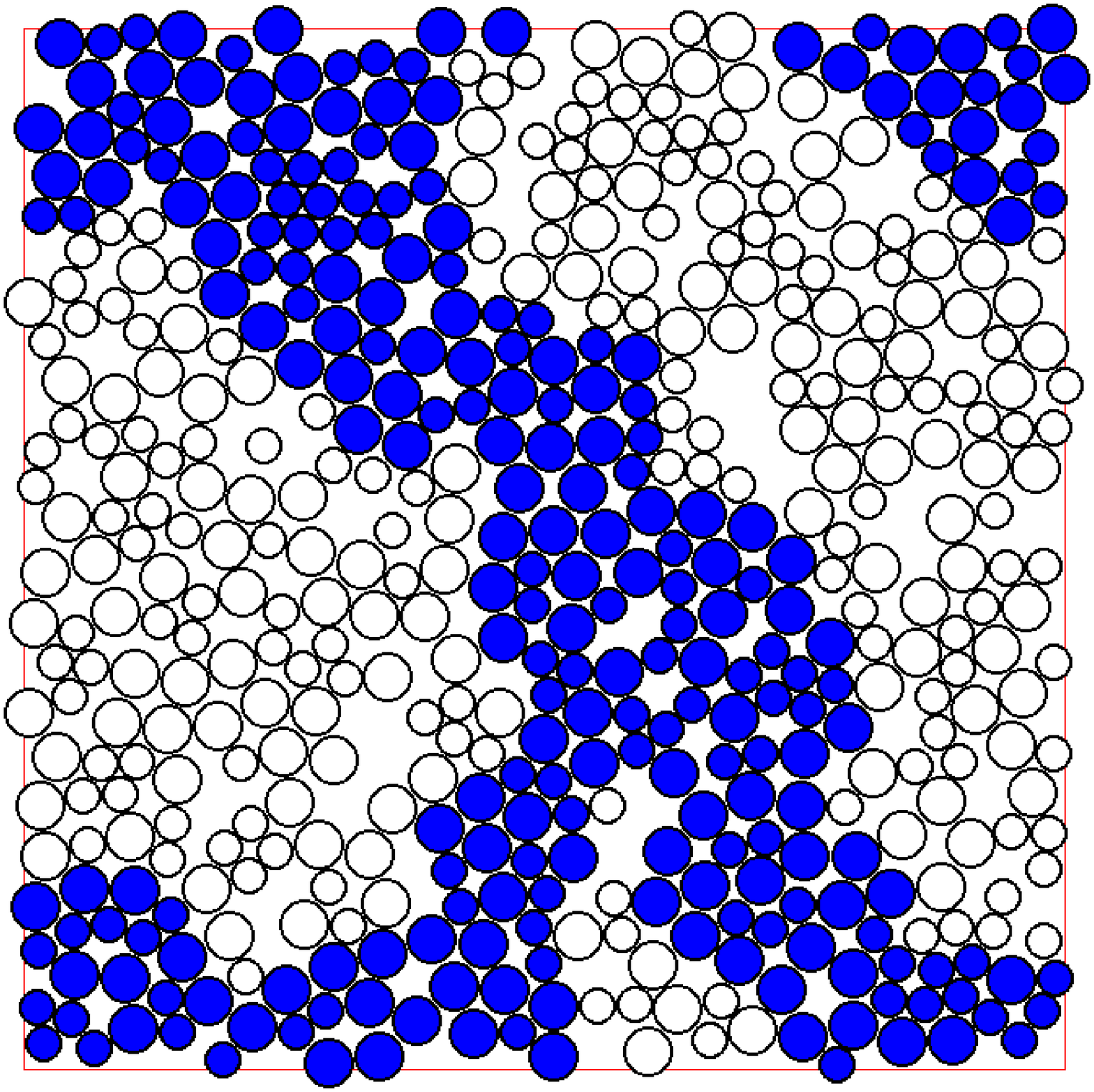}}
}
&
\mbox{
\scalebox{0.16}{\includegraphics{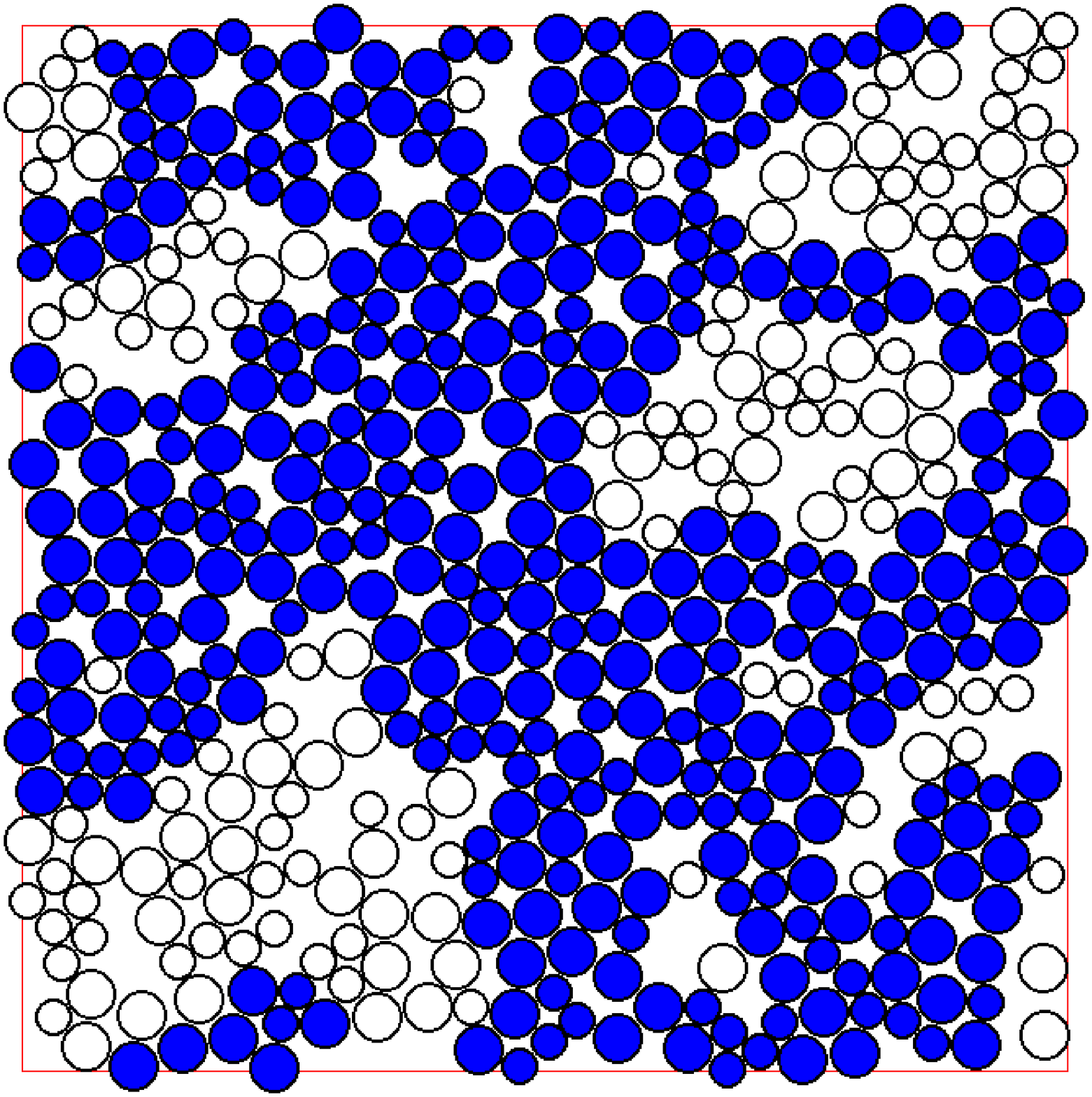}}
} 
\\
{\bf (d)} & {\bf (e)} \\
\mbox{
\scalebox{0.16}{\includegraphics{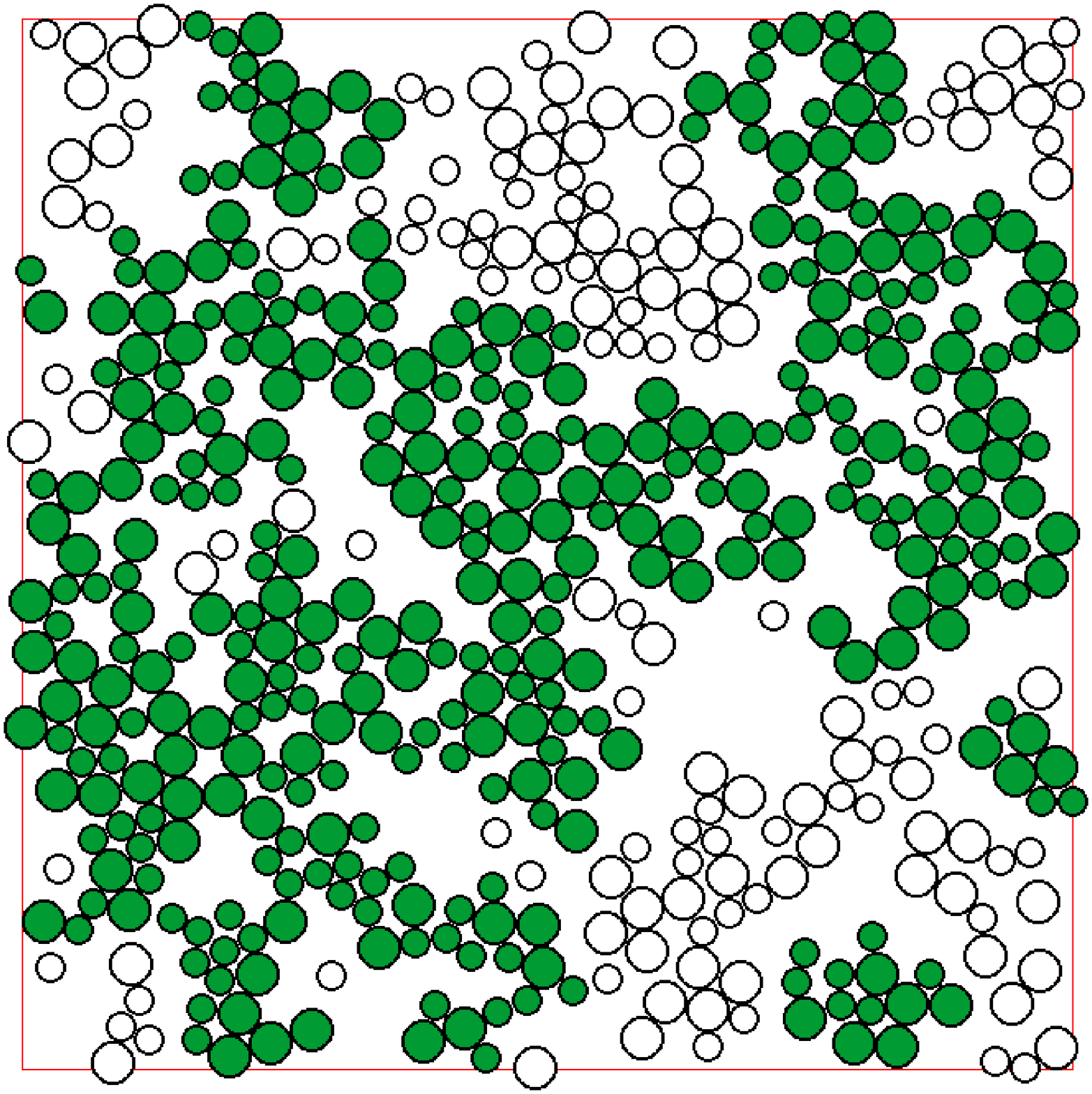}}
}
&
\mbox{
\scalebox{0.16}{\includegraphics{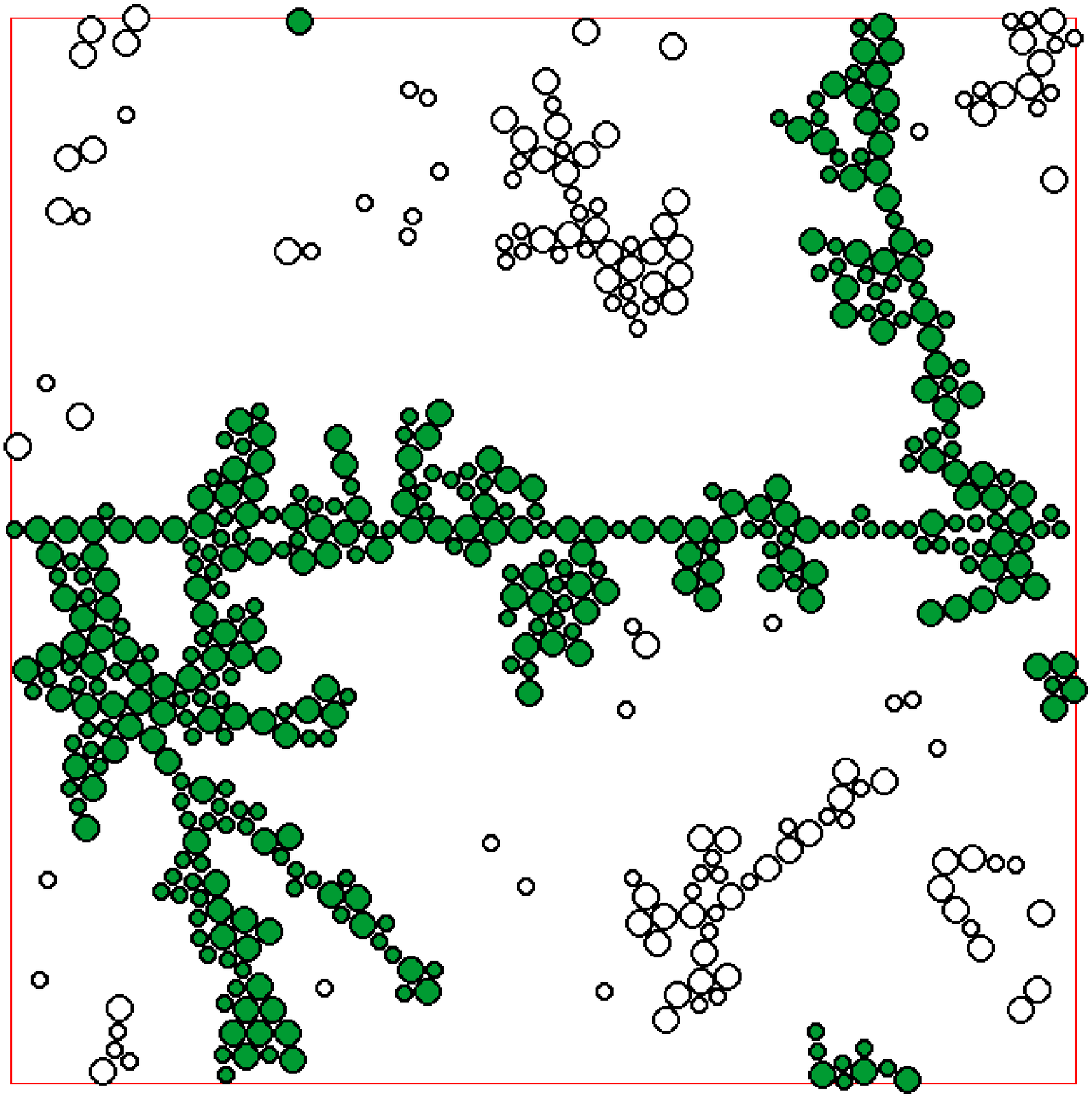}}
}
\\
\end{tabular}
\vspace{-0.1 in}
\caption{ \label{presfig} (Color online).  {\bf (a)} The pressure $P$ vs. $\phi$
for two different compression procedures, which both begin at $P=0$
and $\phi=0.2$.  Red crosses correspond to the jamming path, where the
system is compressed until it becomes jammed at the onset of rigidity percolation
 $\phi_\mathrm{R}$.  Black circles correspond to the connectivity
percolation path, where the system is compressed to the onset of connectivity
percolation $\phi_\mathrm{P}$ and then decompressed.  The
inset shows the central force law used in our simulations.  {\bf (b)}
and {\bf (c)} Screenshots at $\phi_\mathrm{R}$ for
two different systems.  The blue-colored particles comprise the 
percolating backbone.  {\bf (d)} Screenshot at 
$\phi_\mathrm{P}$, with particles in the
percolating cluster colored green.  {\bf (e)} Screenshot of the system
pictured in (d) after decompression, where the percolating structure
is stretched to its fullest extent.  }
\end{center}
\vspace{-0.28 in}
\end{figure}

\paragraph{Simulation procedure}  
We investigate the quasistatic compression of an attractive
particulate system in two and three dimensions at zero
temperature.  The system consists of $N$ particles interacting via a
pairwise, spherically symmetric potential, with a finite repulsive core
and finite-range attraction.  Simulations begin with a dilute
collection of $N$ spherical particles of diameter $\sigma_i$ randomly
placed in a cubic cell with periodic boundary conditions. 
In each simulation step the
diameters of all particles are increased by a small factor and then
the potential energy is minimized using a conjugate gradient method.
The force $F(r_{ij})$ between a pair of particles $i$ and $j$ depends
on their separation $r_{ij}$ relative to the sum of their radii $\sigma_{ij}=(\sigma_i+\sigma_j)/2$,
and is plotted in the inset of
Fig.~\ref{presfig}(a).  Since the overall scale $Y$ of the force is
irrelevant at zero temperature, there are three independent
parameters: the packing fraction $\phi$ (calculated using the location of the minimum in the force law),
the range of the attraction $\ell$, and the minimum scaled force $-C$.

\paragraph{Mechanical Response}  
At zero temperature, whether a particulate system with attractions is
jammed depends on two important factors: a) does the system
contain a percolating cluster and b) if so, does the percolating
cluster possess any floppy modes?  A floppy mode~\cite{alexander} is
an infinitesimal deformation of the material that does not increase
its energy.  The number of floppy modes is equal to the number of
non-trivial zero eigenvalues of the dynamical matrix for a given set
of particle positions and radii~\cite{tanguy}.  If there are no
non-trivial~\cite{trivial} zero eigenvalues, the potential energy
increases with any deformation and all elastic constants are non-zero.
We refer to this state as ``jammed''.  If there is at least one
non-trivial zero eigenvalue, the system is termed ``unjammed''.  Note
that this is a very strict definition of jamming.

It is impossible to have a jammed system that does not
percolate, which leaves three distinct possibilities for its
mechanical state: (1) nonpercolated, (2)
percolated but unjammed, and (3) jammed.  In repulsive systems only (1)
and (3) occur and they are separated by the first-order jamming
transition at point J.  In the attractive particulate systems
investigated here, all three mechanical states occur.  A second-order
connectivity percolation transition separates (1) and (2), and a
second-order rigidity percolation transition separates (2) and (3).

Properties of the three mechanical states are illustrated by measuring
the pressure $P$.  In Fig.~\ref{presfig}(a) we plot $P$ vs. $\phi$ for two
compression procedures in a $d=2$ system with $N=512$, $C=10^{-2}$, and
$\ell=0$. In the first procedure, marked by red crosses, the system is
compressed until it jams at $\phi_\mathrm{R}$, after which the
pressure becomes non-zero.  At $\phi_\mathrm{R}$ a percolating
backbone forms with no floppy modes, also
known as rigidity percolation~\cite{rigidity2d}.  Screenshots of the
system at $\phi_\mathrm{R}$ are shown in Figs.~\ref{presfig}(b) and
(c).  Note that while there are no floppy modes in the percolating
backbone, there are trivial floppy modes in the system involving
particles not in the backbone~\cite{flopper}.

In the second procedure, marked by black circles, the same initial
system is compressed until a percolating network forms at
$\phi_\mathrm{P}$, as illustrated in Fig.~\ref{presfig}(d).
Then the system is decompressed.  We find that upon
decompression the pressure remains zero until a very low packing
fraction.  At this point the network is stretched to its fullest extent 
with no tension, as illustrated in Fig.~\ref{presfig}(e).  
Further decompression causes tension ($P<0$) until the cluster becomes 
unstable, breaks,  and $P=0$.  The formation of a percolating cluster
gives rise to novel mechanical behavior---infinitesimal deformations
do not alter the potential energy but large deformations do. Thus, the
mechanical response of the percolated but unjammed state cannot be
described by linear elasticity.

The mechanical behavior described above is generic.  For all
compaction procedures we investigated, connectivity
percolation and rigidity percolation are distinct
transitions, with $\phi_\mathrm{P}<\phi_\mathrm{R}$.  This occurs
despite the fact that the mechanical response of the system is highly
history dependent.  Since breaking a contact between particles has an
energy cost, contacts gained upon compression are not necessarily lost upon
decompression and we observe that the
number of contacts per particle $z$ never decreases when $P=0$.
Because contacts are not broken,
clusters of particles grow, or coarsen.  This leads to work-hardening
and lowers $\phi_\mathrm{P}$ and $\phi_\mathrm{R}$.

\paragraph{Connectivity Percolation} 
Connectivity percolation occurs when a
cluster of interacting particles first spans the system.
The transition is characterized by three independent critical
exponents, unless the hyperscaling assumption holds, in which case the
number is two ~\cite{percbook}.  The choice of three
independent exponents from a larger number of measurable exponents
is arbitrary.  We choose to focus on (1) the correlation
length exponent $\nu$ which quantifies the divergence of the linear extent $\xi$ of the largest
non-percolating cluster $\xi
\propto |\phi-\phi_\mathrm{P}|^{-\nu}$, (2) the
cluster-statistics exponent $\tau$ which characterizes the
number $n_s$ of clusters with $s$ particles at the percolation
transition $n_s(\phi_\mathrm{P}) \propto s^{-\tau}$, and (3) the
fractal dimension $D$ at $\phi_\mathrm{P}$.  The hyperscaling relation
$D(\tau-1)=d$ holds if $\xi$ is the only relevant length scale in the system.

The critical exponents of the connectivity percolation transition have
been determined for lattices~\cite{percbook} and for continuum
particulate systems without interparticle
forces~\cite{contperc2d,elam}.  The critical exponents, which are
listed in Table~\ref{perctable}, are identical for lattice and
continuum percolation and do not depend on the lattice type.   
In static particulate
systems, force balance is crucial and has not been included in
previous continuum connectivity percolation studies.  In what
follows we measure the independent connectivity percolation exponents,
test the hyperscaling hypothesis, and compare our results to 
lattice connectivity percolation.

\begin{figure}
\begin{center}
\psfrag{yl1}{\Huge{$\Delta^2$}}
\psfrag{xl1}{\Huge{$N$}}
\psfrag{yl2}{\Huge{$n_s(\phi_\mathrm{P})$}}
\psfrag{xl2}{\Huge{$s$}}
\psfrag{yl3}{\Huge{$N_\mathrm{per}$}}
\psfrag{xl3}{\Huge{$N$}}
\scalebox{0.36}{\includegraphics{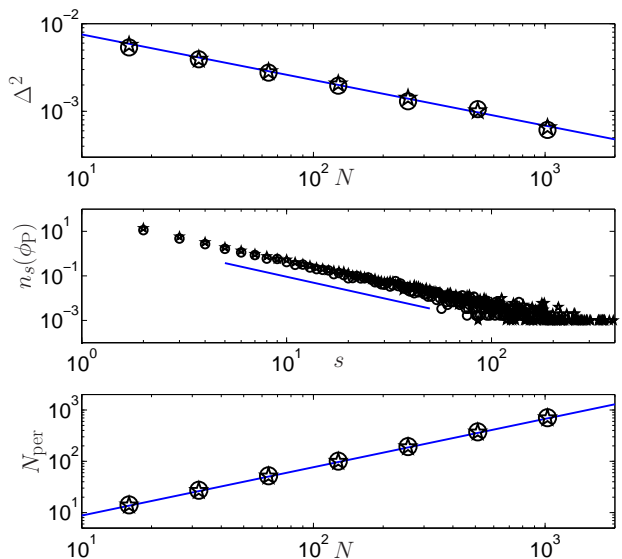}}
\vspace{-0.15 in}
\caption{ \label{percofig} Measurement of the connectivity percolation
exponents in $d=2$.  Circles (stars) correspond to monodisperse
(bidisperse) distributions.  {\bf (a)} The square width of
connectivity percolation thresholds $\Delta^2$ as a function of system
size $N$.  The slope gives $-2/d \nu$.  {\bf (b)} The number of
clusters of size $s$ at $\phi_\mathrm{P}$.
The slope gives $-\tau$.  {\bf (c)} The average number of particles in
the percolation cluster $N_\mathrm{per}$ at $\phi_\mathrm{P}$, as a
function of system size $N$.  The slope gives $D/d$.  The best fit
slopes determined from our analysis are included in each plot.  }
\end{center}
\vspace{-0.3 in}
\end{figure}

In Fig.~\ref{percofig} we show the data used to determine the
connectivity percolation exponents in $d=2$ for both monodisperse and
bidisperse~\cite{endnotebi} systems.  Every data point is
averaged over $5000$ trials, each consisting of an initially random,
dilute state that is subsequently compressed to the percolation
onset at $\phi_\mathrm{P}$.  The correlation length exponent $\nu$ is measured using
the standard finite size scaling analysis~\cite{percbook}, where the root mean square
width $\Delta$ of the $\phi_\mathrm{P}$ distribution is
related to the system size $L$ by $\Delta \propto L^{1/\nu}$.  In
Fig.~\ref{percofig}(a) we plot $\Delta^2$ versus $N$ and observe a
power law behavior with an exponent of
$-2/d\nu$.  This gives $\nu =
1.92 \pm 0.03$, which is larger than the value of $4/3$ for lattice
connectivity percolation.  Next, we determine the exponent $\tau$ by
directly measuring the cluster-size distribution.  In
Fig.~\ref{percofig}(b) we plot $n_s(\phi_\mathrm{P})$ for a system
with $1024$ particles and find that $n_s(\phi_\mathrm{P}) \propto
s^{-\tau}$ with $\tau=2.04 \pm 0.04$.  This is consistent with the
lattice result of $187/91$.  Finally, in Fig.~\ref{percofig}(c) we
plot the average number of particles in the percolating cluster
$N_\mathrm{per}$ as a function of $N$ at $\phi_\mathrm{P}$.  We find
that $N_\mathrm{per} \propto N^{D/d}$, where $D = 1.88 \pm 0.04$ is
the fractal dimension at percolation.  This is also consistent with
the lattice result of $91/48$.

A similar analysis has been applied to $d=3$; the
measured exponents in $d=2$ and $3$ are collected in
Table~\ref{perctable}.  In both cases $D$ and $\tau$ are consistent
with lattice and continuum connectivity
percolation, while $\nu$ is larger than the classical value.  The
hyperscaling relation $D(\tau-1)=d$ is consistent with our results.

\begin{table}
\begin{center}
\begin{tabular}{lccc}
\hline \hline
System & $\nu$ & $\tau$ & $D$ \\ \hline
present work $d=2$ & $1.92 \pm 0.03$ & $2.04 \pm 0.04$  & $1.88 \pm 0.04$ \\
lattice $d=2$~{\cite{percbook}} & $4/3 \, (1.33)$ & $187/91 \, (2.05)$  & $91/48 \, (1.90)$ \\
continuum $d=2$~\cite{contperc2d} & $1.34 \pm 0.02$ & $2.0 \pm 0.1$ & $1.91 \pm 0.04$\\ 
present work $d=3$ & $1.10 \pm 0.03$ & $2.20 \pm 0.03$  & $2.56 \pm 0.06$\\ 
lattice $d=3$~\cite{percbook} & $0.88$ & $2.18$ & $2.53$\\ 
continuum $d=3$~\cite{elam} & $0.94 \pm 0.20$ & $2.22 \pm 0.08$ & $2.21 \pm 0.56$\\ 
\hline \hline
\end{tabular}
\vspace{-0.1 in}
\caption{Critical exponents for connectivity percolation. }
\label{perctable}
\end{center}
\vspace{-0.4 in}
\end{table}

We have tested the robustness of the measured exponents by altering
key parameters in the force law and by varying the preparation
algorithm.  First, we varied the minimum force $C$ over several orders
of magnitude from $10^{-4}$ to $10^{-1}$ and the range of the
attraction $\ell$ from $0$ to $4$.  Second, instead of compressing, we
tested for percolation by quenching to zero temperature at constant
volume.  Third, we redefined the connectivity percolation transition
to require percolation in the $x-$, $y-$ or $z-$directions or all
directions at the same time.  In all of these cases the connectivity
percolation exponents remain the same as those listed in
Table~\ref{perctable}.  We also expect that the measured exponents do
not change if the system is allowed to coarsen from successive
decompressions and compressions or other methods.  If the system
coarsens before it percolates, the length scale over which structure
is affected has a finite extent (equal to the size of the largest
cluster after coarsening).  Near percolation $\xi$ is much larger than
this length scale and coarsening is irrelevant.  However, we
were not able to directly verify this conjecture because of
insufficient system size.

Based on these, considerations we conclude that the
measured exponents (shown in Table~\ref{perctable}) are
universal for attractive, frictionless
particulate systems.  Thus we have identified a new universality class---connectivity
percolation with force balance---that is
different from connectivity percolation in the absence of forces.
The universality class changes abruptly when $Y>0$
(emergence of excluded volume) and $C>0$ (emergence of attraction).

\paragraph{Rigidity Percolation}
The rigidity percolation transition at $\phi_\mathrm{R}$, which
separates the percolated-unjammed and jammed states, occurs
when a set of particles with no floppy modes has percolated.  
We refer to this set as the backbone.  In simulations we
identify backbone particles as those with at least one non-zero
contact force (stressed particles).  Below $\phi_\mathrm{R}$ we always
observe that no particles are stressed and above $\phi_\mathrm{R}$ 
a percolating cluster of stressed particles with no
floppy modes exists.  Figs.~\ref{presfig}(b) and (c) are screenshots of two
systems at rigidity percolation.  We
have also used $P>0$ to define the onset of rigidity percolation, and
obtain identical results~\cite{rigendnote}.

Similar to connectivity percolation, there are three
independent critical exponents in rigidity percolation, and we choose
$\xi$, $\tau$ and $D_b$ (backbone fractal dimension) to 
characterize our system.  Numerical values of
these exponents are listed in Table~\ref{rigperctable} along with the
corresponding lattice results.  Rigidity percolation on a lattice has been studied in
$d=2$~\cite{rigidity2d} and $d=3$~\cite{thorpe}.  In the $d=2$ lattice studies the transition is second-order but only  
the exponents $D_b$ and $\nu$ have been measured. Our measurement of $\nu$ 
differs from the lattice value, while $D_b$ remains the same.
In the $d=3$ lattice studies, the order of
the transition depends on lattice type, which makes comparison with
our off-lattice results difficult.
Hyperscaling is consistent with our results both in $d=2$ and $3$.

We have also measured the fractal dimension $D$ of the
entire percolating cluster, and we find that $D=d$.
Therefore, while the rigidity
percolation transition is second order in the backbone particles, it
might appear to be a discontinuous first-order transition if the
entire percolating cluster is mistaken for the rigid percolating
cluster.

\begin{table}
\begin{center}
\begin{tabular}{lcccc}
\hline \hline
System & $\nu$ & $\tau$ & $D_b$ \\ \hline
present work $d=2$ & $3.12 \pm 0.02$ & $2.08 \pm 0.06$ & $1.77 \pm 0.05$\\
lattice $d=2$~\cite{rigidity2d} & $1.17 \pm 0.02$ & --  & $1.78 \pm 0.02$ \\
present work $d=3$ & $0.92 \pm 0.01$ & $2.31 \pm 0.18$ & $2.58 \pm 0.16$ \\
\hline \hline
\end{tabular}
\vspace{-0.1 in}
\caption{Critical exponents for rigidity percolation.}
\label{rigperctable}
\end{center}
\vspace{-0.4 in}
\end{table}

\paragraph{Extension to finite temperature} 
In this letter, we have considered attractive particulate
systems at zero temperature.  We are currently investigating the effects
of thermal fluctuations.  When $T>0$, contact formation due to particle
or cluster diffusion and contact breaking due to thermal activation
become relevant.  If $T$ is much smaller than the energy needed to
break a contact and the diffusion length scale is small, we expect the
$T=0$ results to hold.  This has been verified for connectivity
percolation in our preliminary studies of these systems undergoing
Brownian dynamics.

Experimental verification of our
predictions can also be found in the literature.  First, in
Ref.~\cite{gel1}, experiments on colloidal gels demonstrate that the
connectivity percolation transition (associated with gelation) occurs
at a lower density than the rigidity percolation transition
(associated with arrested dynamics at all length scales).  This is
consistent with our observation that connectivity and rigidity
percolation occur at different densities.  Second, the critical
exponents for gelation have been measured as $\nu = 1.17
\pm 0.10$~\cite{gel2}, $\tau = 2.15 \pm 0.05$~\cite{gel2p5} in
colloidal suspensions, and $\nu=1.22 \pm 0.16$, $\tau=2.26 \pm 0.08$
in silica gels~\cite{gel3,gel4}.  These values are consistent with the
$d=3$ connectivity percolation exponents measured here for attractive
particulate systems, but are inconsistent with the
connectivity percolation exponents that do not include interparticle
forces.

Financial support from NSF grant numbers CTS-0348175 (GL,JB) and
DMR-0448838 (GL,CSO) is gratefully acknowledged.  We also thank Yale's
High Performance Computing Center for computing time.


\vspace{-0.2 in}
\begin{thebibliography}{100}
\vspace{-0.2 in}

\bibitem{liunagel} A. J. Liu and S. R. Nagel, Nature {\bf 396}, 21 (1998);


\bibitem{zerotemprep} H. A. Makse, D. L. Johnson and L. M. Schwartz, Phys. Rev. Lett. {\bf 84}, 4160 (2000); 
C. S. O'Hern, S. A. Langer, A. J. Liu and S. R. Nagel, Phys. Rev. Lett. {\bf 86}, 111 (2001);
C. S. O'Hern, L. E. Silbert, A. J. Liu and S. R. Nagel, Phys. Rev. E {\bf 68}, 011306 (2003);
H. P. Zhang and H. A. Makse, Phys. Rev. E {\bf 72}, 011301 (2005).
T. S. Majmudar, M. Sperl, S. Luding and R. P. Behringer, Phys. Rev. Lett. {\bf 98}, 058001 (2007).

\bibitem{dynfromjam}
N. Xu, M. Wyart, A. J. Liu and S. R. Nagel, Phys. Rev. Lett. {\bf 98}, 175502 (2007).

\bibitem{attglass}
T. Eckert and E. Bartsch, Phys. Rev. Lett. {\bf 89}, 125701 (2002);
K. N. Pham {\em et. al.}, Science {\bf 296}, 104 (2002).

\bibitem{athermalatt} 
J. M. Valverde, M. A. S. Quintanilla and A. Castellanos, Phys. Rev. Lett. {\bf 92}, 258303 (2004);
V. Richefeu, M. S. El Youssoufi and F. Radjai, Phys. Rev. E {\bf 73}, 051304 (2006);
F. A. Gilabert, J.-N. Roux and A. Castellanos, Phys. Rev. E {\bf 75}, 011303 (2007). 

\bibitem{weitz} V. Trappe, V. Prasad, L. Cipelletti, P. N. Segre and D. A. Weitz, Nature {\bf 411}, 772 (2001).

\bibitem{alexander} S. Alexander, Phys. Rep. {\bf 296}, 65 (1998).

\bibitem{tanguy}
A. Tanguy, J. P. Wittmer, F. Leonforte and J.-L. Barrat, Phys. Rev. B {\bf 66}, 174205 (2002).

\bibitem{trivial}  With periodic boundary conditions there are at least $d$ trivial zero eigenvalues, associated
with translations of the entire system.  In repulsive systems rattlers (particles with no contacts) also introduce trivial zero eigenvalues.  

\bibitem{flopper} ``Floppers'', which can involve many particles, are the generalization of rattlers in purely repulsive systems.

\bibitem{percbook} D. Stauffer, {\em Introduction to Percolation Theory} (Taylor and Francis, London, 1985).

\bibitem{contperc2d} E. T. Gawlinski and H. E. Stanley, J. Phys. A:  Math. Gen. {\bf 14}, L291 (1981).

\bibitem{elam} W. T. Elam, A. R. Kerstein and J. J. Rehr, Phys. Rev. Lett. {\bf 52}, 1516 (1984).

\bibitem{endnotebi} In bidisperse systems half of the particles are large and half are small, with diameter ratio $1.4$.

\bibitem{rigendnote} Above $\phi_\mathrm{R}$,
if a backbone exists then $P>0$.  We
have not been able to prove that $P>0$ implies the existence of a backbone since floppy modes might
exist that are not excited by bulk compression.  Our numerical
results show that $P>0$ is equivalent to the existence of a backbone.

\bibitem{rigidity2d} D. J. Jacobs and M. F. Thorpe, Phys. Rev. Lett. {\bf 75}, 4051 (1995); C. Moukarzel and P. M. Duxbury, Phys. Rev. Lett. {\bf 75}, 4055 (1995).


\bibitem{thorpe} M. V. Chubynsky and M. F. Thorpe, to appear in Phys. Rev. 
E. (2007).

\bibitem{gel1} F. Mallamace, P. Gambadauro, N. Micali, P. Tartaglia, C. Liao and S.-H. Chen, Phys. Rev. Lett. {\bf 84}, 5431 (2000).

\bibitem{gel2} G. Dietler, C. Aubert, D. S. Cannell and P. Wiltzius, Phys. Rev. Lett. {\bf 57}, 3117 (1986).

\bibitem{gel2p5}
A. D. Dinsmore and D. A. Weitz, J. Phys.: Condens. Matter {\bf 14}, 7581 (2002).

\bibitem{gel3} J. E. Martin, J. Wilcoxon and D. Adolf, Phys. Rev. . {\bf 36}, 1803 (1987).

\bibitem{gel4}   $\nu$ is the unswelled value~\cite{gel3} and $\tau$ was inferred from measurements of $\gamma$ and the hyperscaling relation.

\end{thebibliography}
\end{document}